\documentclass[twocolumn,english,showpacs]{revtex4}
\usepackage{babel,amsmath,amssymb,dcolumn}
\usepackage[dvips]{graphics}

\def\imo{i}
\begin{document}

\title{Stability and quasinormal modes of the massive scalar field around Kerr black holes}

\author{R.A. Konoplya}
\email{konoplya@fma.if.usp.br}
\author{A. V. Zhidenko}
\email{zhidenko@fma.if.usp.br}
\affiliation{Instituto de F\'{\i}sica, Universidade de S\~{a}o Paulo \\
C.P. 66318, 05315-970, S\~{a}o Paulo-SP, Brazil}

\pacs{04.30.Nk,04.50.+h}

\begin{abstract}
We find quasinormal spectrum of the massive scalar field in the
background of the Kerr black holes. We show that all found modes
are damped under the quasinormal modes boundary conditions when $\mu
M$ is not large, thereby implying stability of the massive scalar field. This
complements the region of stability determined by the Beyer inequality for
large masses of the field. We show that, similar to the case of a non-rotating black holes, the
massive term of the scalar field does not contribute in the regime
of high damping. Thereby, the high damping asymptotic should be
the same as for the massless scalar field.
\end{abstract}

\maketitle

\section{Introduction}

Scalar field in a black hole backgound is a subject of active
investigation \cite{scalar-field}, yet the massive scalar field
has been investigated only in very few studies as to its quasinormal
spectrum \cite{2},\cite{Cimone}, late-time behavior \cite{Burko} and scattaring
properties \cite{Park}. Behavior of the massive scalar field in a
black hole background is quite different from that of the massless
one in many aspects: first, it demonstrates the so-called
superradiant instability \cite{superbomb}, which is absent for
massless fields. This super-radiant  regime happens provided the
real oscillation frequency satisfies the inequality $| Re
\omega | < m a/(2M r_{+})$, where $m$ is the azimutal number of the mode,
and $M$ is the mass of the black hole. Another research, indicating
the instability of massive scalars was that by
Detweiler \cite{Detweiler} where slowly growing instability was
suggested for values $\mu M << 1$, i.e. in the regime, when
particle Compton wavelength is much larger then the size of the
black hole ($\mu$ is the mass of the scalar field). Second, it may cause
existence of infinitely long-living modes called quasi-resonances
\cite{3}. Finally, at asymptotically late times the massive fields
show universal behavior independent on spin of the field
\cite{K-M}. Note, that scalar field with the mass term can be also
interpreted as a self-interacting scalar field within regime of
small perturbations
\cite{Hod}. One more recent application of dynamic of the massive scalar field  in a
black hole backgound comes from different brane-world theories
where massless fields can be considered as effectively massive
fields on the brane.

Yet, the strongest motivation to make an extensive search for
quasinormal modes of massive scalar field is that until now the
stability for massive scalar field perturbations in the Kerr
background is not studied under non-super-radiant (quasinormal)
boundary conditions. Usually, stability of spherically symmetric
black holes, like Schwarzschild or Reissner-Nordrstrom ones, can be
relatively easily proved by using positivity of the conserved
energy. On the contrary to the Schwarzschild case, Kerr black holes have
negative energy density inside the ergo-sphere. Therefore, fields
can grow in parts of the space-time, leading to instability. 
The only result on stability of massive scalar field in Kerr
space-time, we are aware of, is that of Beyer
\cite{Beyer}, where it was proved that the modes of field with the
mass obeying the inequality
\begin{eqnarray}
\mu_{0} \geq \frac{| m | a}{2 M r_{+}} \sqrt{1 + \frac{2
M}{r_{+}} + \frac{a^{2}}{r_{+}^{2}}},
\end{eqnarray}
are stable. Note, that the inequality found by Beyer is not
compatible with conditions for super-radiance, so it does not
contradict super-radiant instability of massive fields. In
addition, the above inequality does not mean that fields of lower
mass are unstable, and Beyer suggested the ways to lower the
inequality
\cite{Beyer}. If such an instability exists one could find an
accurate threshold value of field mass $\mu$ when the instability
begins. This could be done by an extensive computing of the
quasinormal modes and by finding of the growing ones. Some low
laying WKB quasinormal modes were found by Simone and Will
\cite{Cimone}, yet neither fundamental modes ($\ell = n = 0$, $\ell$ is multi-pole number), nor
modes with  $n \geq \ell$ can be found with trustable accuracy by
WKB approach \cite{13,13a,13b}, even despite very good accuracy of WKB
method for low overtones \cite{WKBapply}. Therefore, being limited by WKB
accuracy, one cannot judge about stability and high damping
asymptotic in this case. Recently the quasinormal damping and late
time regimes in time domain were investigated by L. Burko and G. Khanna in
\cite{Burko}, and no instability was observed in the considered range
of values  of field mass $\mu$ \cite{Burko}.

Because of the above reasons, in this paper we used the accurate
Frobenius method to make an extensive search for quasinormal modes
of the massive scalar field around Kerr black holes. The paper
organized as follows: Sec. II introduces the main formulas for the
wave equations and the Frobenius procedure. Sec. III gives
numerical data for found quasinormal modes, and Sec. IV discusses
the obtained results and the possibility for (un)stability.

\section{Basic equations}

The background metric in the Boyer-Lindquist coordinates $t$, $r$,
$\theta$, $\phi$ has the form:
\begin{eqnarray}
ds^2=\left(1-\frac{2 M r}{\Sigma}\right)dt^2+\frac{4 a M r
\sin^2\theta}{\Sigma
}dtd\phi-\frac{\Sigma}{\Delta}dr^2&&\nonumber\\-\Sigma
d\theta^2-\left(r^2+a^2+\frac{2 M a^2 r
\sin^2\theta}{\Sigma}\right)\sin^2\theta d\phi^2,&&\label{KerrLE}
\end{eqnarray}
where, $M$ is the mass of the black hole,
$$\Delta=r^2+a^2-2Mr, $$
and
$$\Sigma=r^2+a^2\cos^2\theta.$$
The massive scalar field obeys the equation:
\begin{eqnarray}
\Box \Phi + \mu^2 \Phi = 0.
\end{eqnarray}
The radial wave equation ($\Phi = e^{-\imo\omega t}e^{im\phi} R(r) S(\phi, \theta)$) is well-known:
\begin{equation}\label{radialeq}
\frac{d}{dr}\left(\Delta\frac{dR(r)}{dr}\right)+\left(\frac{K^2}{\Delta}-\lambda-\mu^2r^2\right)R(r)=0,
\end{equation}
where 
\begin{eqnarray}
K= \omega(r^{2} + a^{2}) - a m. 
\end{eqnarray}

The separation constant $\lambda$ satisfies the equation
\begin{eqnarray}
&&\frac{1}{\sin\theta}\frac{d}{d\theta}\left(\sin\theta\frac{dS(\theta)}{d\theta}\right)+
\Biggr(-\frac{m^2}{\sin^2\theta}\label{angulareq}\\\nonumber
&&-a^2\omega^2\sin^2\theta
-a^2\mu^2\cos^2\theta+2am\omega+\lambda\Biggr)S(\theta)=0,
\end{eqnarray}
and can be found numerically for any given $\omega$ \cite{Suzuki:1998vy}.

Since $\lambda(\omega)$ can be obtained numerically, we are able to solve the
equation for the radial part (\ref{radialeq}), using the continued fraction method
\cite{Leaver:1985ax}. In order to satisfy the quasinormal mode boundary
conditions,  which corresponds to purely in-going waves at the
event horizon and purely outgoing waves at spatial infinity, we require
that
\begin{eqnarray}\nonumber
R(r\rightarrow\infty) &\propto& \exp(\imo\chi
r)r^{\imo\sigma-1},\quad\sigma=\frac{(\omega^2+\chi^2)(r_+^2+a^2)}{2\chi
r_+},\\
\nonumber
R(r\rightarrow r_+)&\propto& (r-r_+)^{-\imo\alpha},\quad\alpha=\frac{\omega r_+ (r_+^2+a^2)-mar_+}{r_+^2-a^2},
\end{eqnarray}
where $r_+$ marks the event horizon:
$$r_+ = M + \sqrt{M^{2} - a^{2}},$$
and $$\chi=\pm\sqrt{\omega^2-\mu^2}.$$ The sign for  $\chi$ should  be chosen in
order to remain in the same complex quadrant as $\omega$.

The equation (\ref{radialeq}) has three singularity points: the
spatial infinity ($r=\infty$), the event horizon ($r=r_+$) and the
internal horizon ($r=r_-=a^2/r_+<r_+$). Thefore the appropriate
Frobenius series has the form:
\begin{eqnarray}
R(r)=\exp(\imo\chi r)\left(\frac{r}{r_+}-\frac{a^2}{r_+^2}\right)^{\imo\sigma-1}\left(\frac{rr_+-r_+^2}{rr_+-a^2}\right)^{-\imo\alpha}\times\nonumber\\\times\sum_{i=0}^\infty a_i\left(\frac{rr_+-r_+^2}{rr_+-a^2}\right)^i.
\end{eqnarray}

And the coefficients satisfy the three-term recurrence relation:
\begin{equation}
\alpha_na_{n+1}+\beta_na_n+\gamma_na_{n-1}=0, \qquad n\geq0, \quad\gamma_0=0,
\end{equation}
where $\alpha_n$, $\beta_n$, $\gamma_n$ can be found in analytical
form. We do not present them here because they have rather combersome
form.

By comparing the ratio of the series coefficients
\begin{eqnarray}%
\frac{a_{n+1}}{a_n}&=&\frac{\gamma_{n}}{\alpha_n}\frac{\alpha_{n-1}}{\beta_{n-1}
-\frac{\alpha_{n-2}\gamma_{n-1}}{\beta_{n-2}-\alpha_{n-3}\gamma_{n-2}/\ldots}}-\frac{\beta_n}{\alpha_n},\nonumber\\
\label{ratio}\frac{a_{n+1}}{a_n}&=&-\frac{\gamma_{n+1}}{\beta_{n+1}-\frac{\alpha_{n+1}\gamma_{n+2}}{\beta_{n+2}-\alpha_{n+2}\gamma_{n+3}/\ldots}},
\end{eqnarray}%
we obtain an equation with a convergent \emph{infinite continued
  fraction} on its right side:
\begin{eqnarray}\label{continued_fraction} \beta_n-\frac{\alpha_{n-1}\gamma_{n}}{\beta_{n-1}
-\frac{\alpha_{n-2}\gamma_{n-1}}{\beta_{n-2}-\alpha_{n-3}\gamma_{n-2}/\ldots}}=\qquad\\\nonumber
\qquad=\frac{\alpha_n\gamma_{n+1}}{\beta_{n+1}-\frac{\alpha_{n+1}\gamma_{n+2}}{\beta_{n+2}-\alpha_{n+2}\gamma_{n+3}/\ldots}},
\end{eqnarray}%
that can be solved numerically by minimizing the absolute value of the
difference between its left and right sides. The equation
(\ref{continued_fraction}) has infinite number of roots, but the most
stable root depends on $n$. Generally the larger number $n$
corresponds to the larger imaginary part of the root $\omega$.

Note, that the case under consideration allows to use the Nollert
procedure \cite{Nollert}, in order to improve convergence  of the
infinite continued fraction, that is useful for searching roots
with a very large imaginary part.

\section{Numerical results}

Since we are limited by Kaluza-Klein equation, we do not take into
consideration the effect of the mass of the scalar field onto the
black hole. Thereby, we can consider only small values of mass of
the field $\mu$, in comparison with the mass of the black hole $M$,
i.e. $\mu M << 1$. Yet, as large values of $\mu$ are possible in
brane-world theories, we shall touch this case as well.

The quasinormal modes of the Kerr black hole depend on several
parameters: the multi-pole number $\ell$, the overtone number $n$,
the azumutal number $m$, and on the black hole and scalar field
parameters: $M$, $a$, $\mu$. Note that $m$ can take values
$-\ell,-\ell+1, ..., \ell-1, \ell$. From here and on we measured 
$\mu$ and $a$ in units of black hole mass $M$.

Let us look at tables I-III: the damping rate determined by the
imaginary part of the $\omega$ is decreasing considerably with
increasing of mass $\mu$. While the real oscillation frequency is
slightly increased with grwoing of $\mu$. This happens also for Schwarzschild
black hole \cite{2}. Note, that it is the fundamental mode
$n=0$ that contributes mainly into the ringing signal.

To check the correctness of the applied procedure we considered
particular cases of massless scalar field in the Kerr background
and of the massive field in the Schwarzschild background. For these two cases the quasinormal modes are well-known for example
from the papers \cite{particular},  \cite{2}, \cite{Cimone}, \cite{3}.
One can see, for instance in the table III that QNMs in case of $\mu =
0$, coincide with corresponding data in of \cite{particular},
\cite{2}, \cite{Cimone}. Also data in tables I-IV for $a=0$ coincide
with modes found in \cite{3}.

\begin{widetext}

\begin{table}
\caption{Values of the quasinormal frequencies for the fundamental
  mode $n = 0$, $\ell = m = 0$ for different values of mass $\mu$, and rotation $a$.}
\label{GB_QNM_1}
\begin{ruledtabular}
\begin{tabular}{ccccccc}
\multicolumn{1}{l}{$\ell = m = 0$} &
\multicolumn{2}{c}{$\mu=0.1$}&
\multicolumn{2}{c}{$\mu=0.2$}&
\multicolumn{2}{c}{$\mu=0.3$}\\
a & Re($\omega_0$) & -Im($\omega_0$)  & Re($\omega_0$) & -Im($\omega_0$) & Re($\omega_0$) & -Im($\omega_0$)
\\
\hline
\\
0 &   0.112361 & 0.096823   & 0.116207 & 0.075358  & 0.122060   & 0.043605  \\
0.1 & 0.112447 & 0.096741   & 0.116311 & 0.075312  & 0.122442   & 0.043495  \\
0.2 & 0.112707 & 0.096490   & 0.116625 & 0.075167  & 0.123554   & 0.043800 \\
0.3 & 0.113140 & 0.096050  &  0.117153 & 0.074910  & 0.123993   & 0.044905    \\
0.4 & 0.113747 & 0.095387   & 0.117903 & 0.074511  & 0.123805 &   0.044021  \\
0.5 & 0.114521 & 0.094442  &  0.118883 & 0.073988  & 0.126101 &   0.045025   \\
0.6 &  0.115442 &  0.093120  &  0.120093 & 0.073056  & 0.128069  & 0.044110  \\
0.7 &  0.116442 &  0.091261  &  0.121514 & 0.071782  & 0.130044 &  0.044193   \\
0.8 &  0.117301 &  0.088592 &   0.123033 &  0.069849  & 0.129843 & 0.043152  \\
0.9 &  0.117181 &  0.084742  &  0.124136 &  0.066839  & 0.135784  & 0.041884 \\
0.95 & 0.115848 &  0.082697  &  0.123910 & 0.064981  &  0.136616  & 0.042938   \\
0.99 & 0.114478 &  0.082285  &  0.123197  & 0.064010   & 0.134994 & 0.040595 \\
0.995 &0.114388  & 0.082240   & 0.123144  & 0.063928   & 0.136560 & 0.040180 \\
\end{tabular}
\end{ruledtabular}
\end{table}
\end{widetext}
\narrowtext

As the number of the overtone is increased, the QN frequency of
the massive field  becomes closer to that of the massless field. 
For the case of Schwarzschild black hole, we proved in \cite{3}, both
analytically and numerically, that this tendency of QN modes to
diminsh difference stipulated by mass $\mu$, ends with a universal
high damping asymptotic, which is independent on the mass of the
field. Thus, the massive term $\mu$ does not contribute to high
overtone behavior of Schwarzschild black holes.
We shall show numerically that the same is true for Kerr black holes.
Indeed, from Table II, one can see that already for relatively small
overtone number, the difference between quasinormal modes with
different values of mass $\mu$ becomes negligible. This is important 
for us, because now, we are in position to say that it is enough to 
study low overtone behavior to judge  about stability of Kerr case
within our numerical approach.

\begin{table}
\caption{QNMs for $\mu =0.3$ and $\mu=0.1$, $a=0.5$, $\ell=m=1$.}
\begin{ruledtabular}
\begin{tabular}{ccc}
$n$ & $\mu=0.3$ & $\mu=0.1$ 
\\
\hline
\\
1 & 0.330086 - 0.267718 i   & 0.325510 - 0.288297 i  \\
2 & 0.292386 - 0.489967 i  & 0.296315 - 0.501634 i  \\ 
3 & 0.266009 - 0.720273 i  & 0.270472 - 0.726246 i  \\
4 & 0.245565 - 0.952427 i   & 0.249182 - 0.955699 i  \\
5 & 0.228436 - 1.184966 i  & 0.231248 - 1.186916 i  \\
10 & 0.164599 - 2.343964 i & 0.165873 - 2.344194 i  \\
15 & 0.117999 - 3.499102 i &0.118599 - 3.499165 i   \\
\end{tabular}
\end{ruledtabular}
\end{table}

From Tables I-V and figures (1-2), we see that all found modes for different
values of parameters $\ell$ and $m$ are damped in the regime of
small $\mu$. Larger $\mu$ (see Fig. 3) leads to considerable decreasing of the
damping rate and finally to the same infinitely long living modes
(i.e. modes with infinitesimaly small imaginary part) found in
\cite{3} for Schwarzschild black holes and called {\it
quasi-resonances}. Therefore,  when increasing $\mu$, the strict
line in Fig. 3 does not have continuation into negative half-plane. 
We see here that, the larger mass of the field $\mu$ is, the more
low laying modes will change into quasi-resonances. Meanwhile, the rest 
of the spectra consists of ordinary modes (i.e. modes with
finite time of damping). Moreover, at sufficiently high overtone
number, the difference between massive and massless case will be
neglected (see Table II). This behavior mimics the QN behavior  of
non-rotating black holes. Yet, potential for massive scalar field around
non-rotating black holes is definitely positive everywhere outside
black hole, while for Kerr solution, the potential depends upon
complex frequency and such a stightforward analysis is impossible. 
 
Now we are in position to state that the high damping asymptotic for 
quasinormal modes of massive scalar field is the same as for massless
one, i.e. the real part increases with $m$ according to the numerical
law described in \cite{Berti:2004um}.

\begin{figure}
\resizebox{1\linewidth}{!}{\includegraphics*{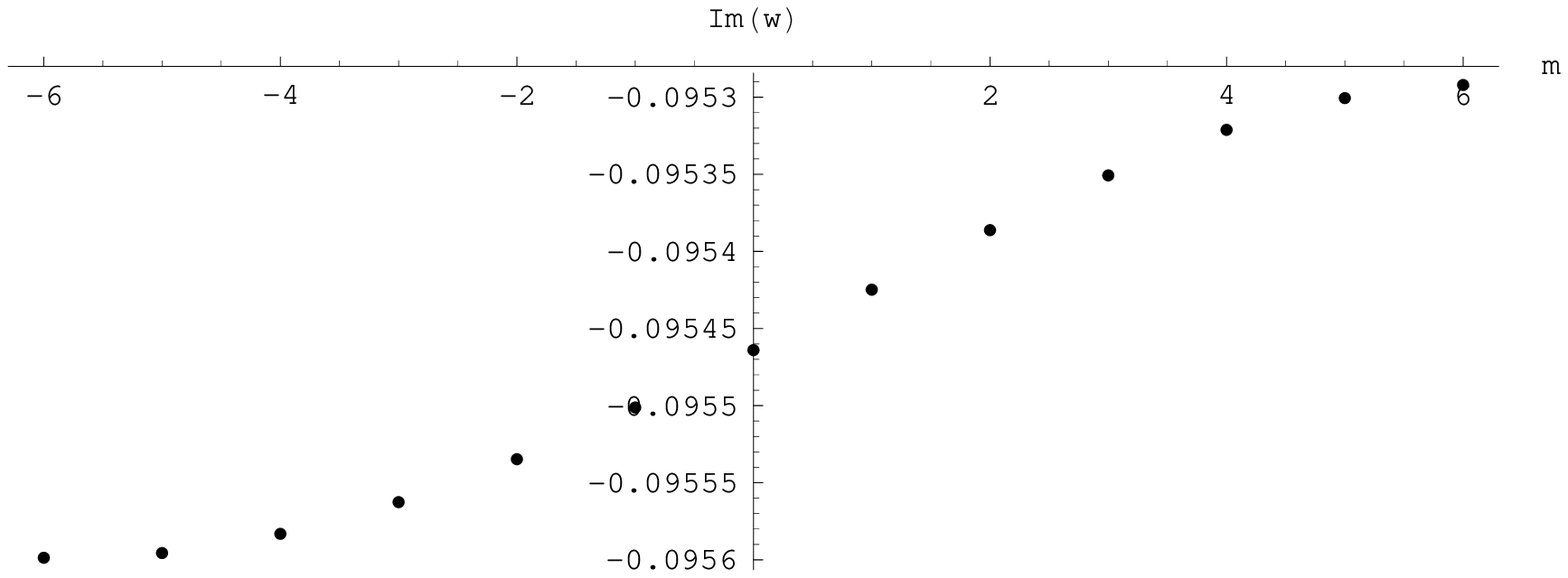}}
\caption{$Im \omega$ as a function of azimutal number $m$ for $\ell =
6$, $a=0.3$, $n=0$, $\mu =0.1$.}
\label{potential1}
\end{figure}
\begin{figure}
\resizebox{1\linewidth}{!}{\includegraphics*{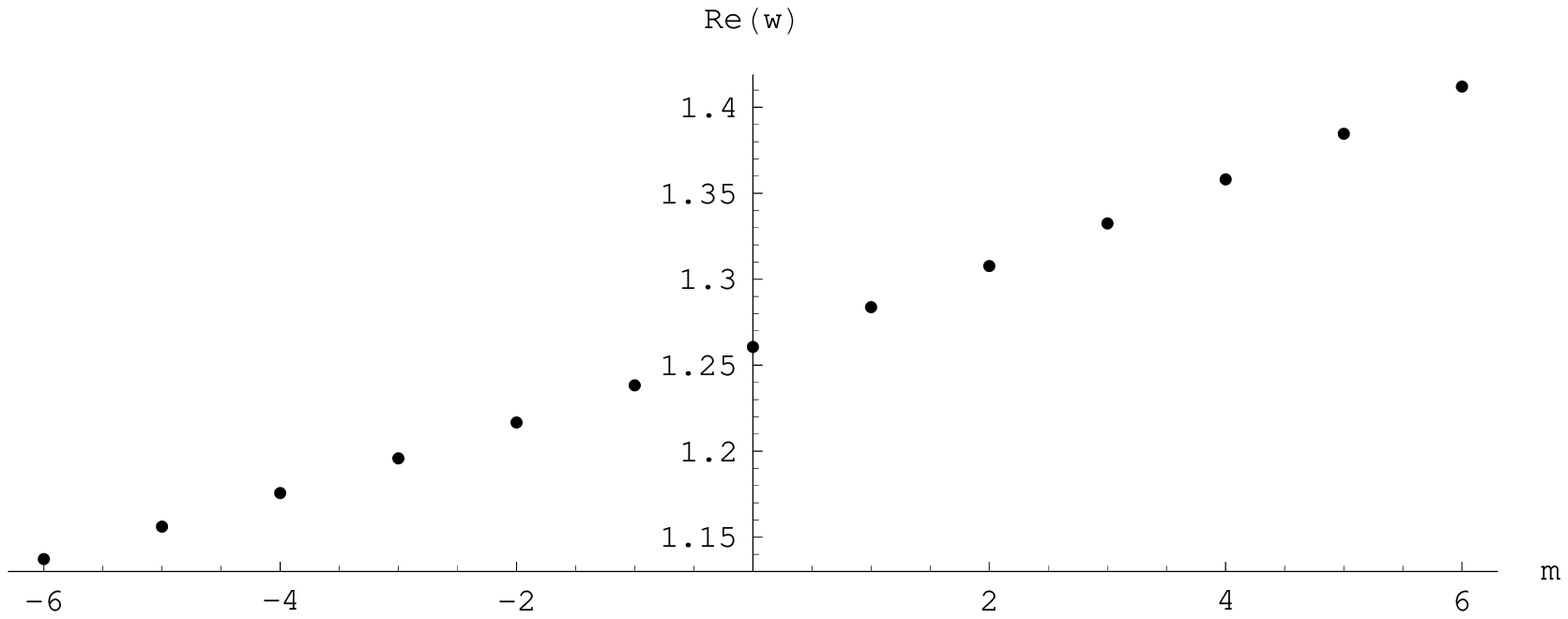}}
\caption{$Im \omega$ as a function of azimutal number $m$ for $\ell =
6$, $a=0.3$, $n=0$, $\mu =0.1$.}
\label{potential1}
\end{figure}

To exclude possible instability we should search for QNMs for
different values of $\ell$ and $m$. As one can see from representative
plots Fig. 1, at large $\ell$ the imaginary part of the
QN frequencies remains bounded and, thereby, does not show any tendency
to instability. The dependence of the real osicillation frequency 
on the azimutal number $m$ can be found on Fig. 3.

After careful investigating of region of small values $\mu M$, 
we conlcude that there are {\it no unstable modes}. As to the case
of intermediate $\mu M$, our extensive search of quasinormal modes 
implies that there are no unstable modes at least as far as  $\mu M$
is not much larger then $1$. We certainly cannot find quasinormal 
modes for any large value of $\mu$. This happens because the numerical
procedure used here requires slow changing of $\mu$, when searching
for new modes, in order not to jump occassionaly into another overtone.
Thereby, considerably large values of $\mu$ can be reached only for 
very long time of computation.

\begin{widetext}

\begin{table}
\caption{Values of the quasinormal frequencies for the fundamental
  mode $n = 0$, $\ell = 1$, $m = 0$ for different values of mass $\mu$, and rotation $a$.}
\label{GB_QNM_2}
\begin{ruledtabular}
\begin{tabular}{ccccccc}
\multicolumn{1}{l}{$\ell = 1; m = 0$} &
\multicolumn{2}{c}{$\mu=0.1$}&
\multicolumn{2}{c}{$\mu=0.2$}&
\multicolumn{2}{c}{$\mu=0.3$}\\
a & Re($\omega_0$) & -Im($\omega_0$)  & Re($\omega_0$) & -Im($\omega_0$) & Re($\omega_0$) & -Im($\omega_0$)
\\
\hline
\\
0 &   0.297416 & 0.094957   &  0.310957 &  0.086593  & 0.333777 &  0.071658  \\
0.1 & 0.297602 & 0.094884  &   0.311128  & 0.086544  & 0.333924 &  0.071646  \\
0.2 & 0.298164 & 0.094661  &   0.311646 &  0.086392  & 0.334366 &  0.071608 \\
0.3 & 0.299116 & 0.094273  &   0.312523 &  0.086124  & 0.335116 &  0.044905    \\
0.4 & 0.300478 & 0.093692   &  0.313780 &  0.085717  & 0.336190  & 0.071408  \\
0.5 & 0.302285 & 0.092873  &   0.315446 &  0.085132  & 0.337622 &  0.711202  \\
0.6 & 0.304579 & 0.091742  &   0.317564 &  0.084308  & 0.339444  & 0.070869   \\
0.7 & 0.307414 & 0.090182  &   0.320185 &  0.083141  & 0.341707 &  0.070335   \\
0.8 & 0.310836 & 0.087989 &    0.323357 &  0.081461  & 0.344465 &  0.069472  \\
0.9 & 0.314815 & 0.084810  &   0.327070 &  0.078967  & 0.347735  & 0.068059 \\
0.95 &0.316919 & 0.082696  &   0.329061  & 0.077284  & 0.347735  & 0.068058  \\
0.99 &0.318576 & 0.080718  &   0.351004 &  0.066054 &  0.351004 &  0.066054 \\
0.995 &0.318779 & 0.080453  &  0.351189 &  0.065918  & 0.351189  & 0.065918 \\
\end{tabular}
\end{ruledtabular}
\end{table}

\end{widetext}

\begin{widetext}

\begin{table}
\caption{Values of the quasinormal frequencies for the fundamental mode $n = 0$, $\ell = 1$, $m = 1$ for different values of mass $\mu$, and rotation $a$.}
\label{GB_QNM_3}
\begin{ruledtabular}
\begin{tabular}{ccccccc}
\multicolumn{1}{l}{$\ell = 1; m = 1$} &
\multicolumn{2}{c}{$\mu=0$}&
\multicolumn{2}{c}{$\mu=0.1$}&
\multicolumn{2}{c}{$\mu=0.2$}\\
a & Re($\omega_0$) & -Im($\omega_0$)  & Re($\omega_0$) & -Im($\omega_0$) & Re($\omega_0$) & -Im($\omega_0$)
\\
\hline
\\
0   & 0.292936 & 0.097660 & 0.297416 & 0.094957 & 0.310957 & 0.086593 \\

0.1 & 0.301045 & 0.097547 & 0.305329 & 0.095029 & 0.318274 & 0.087228 \\

0.2 & 0.310043 & 0.097245 & 0.314119 & 0.094920 & 0.326433 & 0.087709 \\

0.3 & 0.320126 & 0.096691 & 0.323981 & 0.094569 & 0.335621 & 0.087979 \\

0.4 & 0.331567 & 0.095792 & 0.335181 & 0.093883 & 0.346095 & 0.087950 \\

0.5 & 0.344753 & 0.094395 & 0.348105 & 0.092714 & 0.358230 & 0.087478 \\

0.6 & 0.360285 & 0.092243 & 0.363345 & 0.090805 & 0.372594 & 0.086320 \\

0.7 & 0.379159 & 0.088848 & 0.381888 & 0.087678 & 0.390141 & 0.084014 \\

0.8 & 0.403273 & 0.083132 & 0.405606 & 0.082262 & 0.412675 & 0.079526 \\

0.9 & 0.437234 & 0.071848 & 0.439045 & 0.071342 & 0.444549 & 0.069737 \\

0.95& 0.462261 & 0.060091 & 0.463691 & 0.059825 & 0.468050 & 0.058968 \\

0.99& 0.493423 & 0.036712 & 0.494284 & 0.036756 & 0.496939 & 0.036879 \\

0.995&0.490424 & 0.073874 & 0.490872 & 0.073846 & 0.492223 & 0.073740 \\
\end{tabular}
\end{ruledtabular}
\end{table}
\end{widetext}

\begin{widetext}

\begin{table}
\caption{Values of the quasinormal frequencies for higher overtones
for $\mu M = 0.1$, $\ell = 1$, $m = 1$.}

\label{GB_QNM_4}

\begin{ruledtabular}
\begin{tabular}{ccccccc}

\multicolumn{1}{l}{$\ell = 1; m = 1$} &
\multicolumn{2}{c}{$n=1$}&
\multicolumn{2}{c}{$n=2$}&
\multicolumn{2}{c}{$n=3$}\\

a & Re($\omega_0$) & -Im($\omega_0$)  & Re($\omega_0$) & -Im($\omega_0$) & Re($\omega_0$) & -Im($\omega_0$)
\\
\hline
\\

0   & 0.264689 & 0.302851 & 0.228650 & 0.538599 & 0.202558 & 0.787632 \\

0.1 & 0.274260 & 0.301711 & 0.239156 & 0.534670 & 0.212898 & 0.780859 \\

0.2 & 0.284872 & 0.299936 & 0.250883 & 0.529466 & 0.224537 & 0.772046 \\

0.3 & 0.296738 & 0.297332 & 0.264079 & 0.522633 & 0.237749 & 0.760644 \\

0.4 & 0.310140 & 0.293608 & 0.279072 & 0.513640 & 0.252894 & 0.745820 \\

0.5 & 0.325510 & 0.288297 & 0.296315 & 0.501635 & 0.270471 & 0.726246 \\

0.6 & 0.343441 & 0.280603 & 0.316481 & 0.485169 & 0.291244 & 0.699664 \\

0.7 & 0.364932 & 0.269019 & 0.340644 & 0.461505 & 0.316511 & 0.661835 \\

0.8 & 0.391790 & 0.250187 & 0.370780 & 0.424542 & 0.348980 & 0.603557 \\

0.9 & 0.428112 & 0.213833 & 0.411782 & 0.355839 & 0.382103 & 0.644425 \\

0.95& 0.441634 & 0.287111 & 0.431664 & 0.399149 & 0.382103 & 0.644425 \\

0.99& 0.473604 & 0.275655 & 0.468124 & 0.395139 & 0.460787 & 0.576199 \\

0.995&0.480613 & 0.293136 & 0.477664 & 0.381687 & 0.471122 & 0.604783 \\
\end{tabular}
\end{ruledtabular}
\end{table}
\end{widetext}

\begin{figure}
\resizebox{1\linewidth}{!}{\includegraphics*{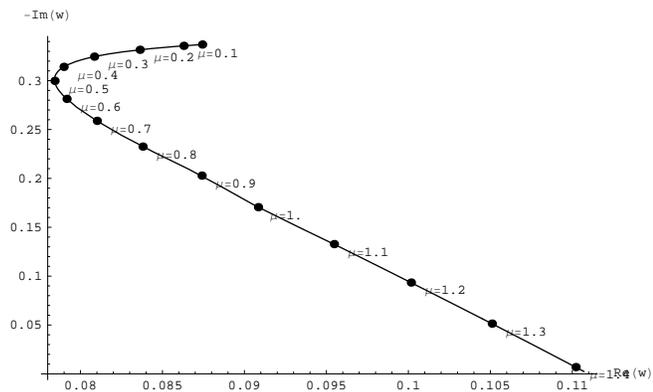}}
\caption{$Re \omega$ vs $Im \omega$ for $n=1$, $a=0.5$,
$\ell = m = 0$ for different values of $\mu$ }
\label{potential1}
\end{figure}

\section{Conclusion}

In the present paper we have considered the massive scalar field
perturbations for the Kerr black holes. Within numerical approach we
managed to compute quasinormal modes for values $\mu M $ not much
larger then $1$. In this regime all found low laying modes are damped.  
We have shown that similar to the Schwrazschild case, the massive 
term does not contribute into the high damping part of the spectrum.
Therefore we can conclude that for not large  values of $\mu M $
there is no instability for massive scalar field under the
quasinormal mode boundary conditions.  From the inequality (1) it follows that
for a given black hole mass $M$ and rotation $a$, for any large $\mu$, there is
sufficiently large $m$, so that inequality is not performed. Therefore
we sould have instability for sufficiently large values of $m$, even
for large $\mu$. Yet, we have whon in Fig. 1, that QN spectrum for larger values of
$m$ does not show any tendency to instability, because the  $Im \omega$
for increasing values of $m$ remains bounded within some region of
negative values corresponding to damping.  
Thefore inequlity (1) found in \cite{Beyer} means stability for fields with
sufficiently large values of $\mu$.

In addition, we have shown that
the qusi-resonances, which exist for the Schwrazschild metric, exist also
for Kerr black holes.

\begin{acknowledgments}
A. Z. and R. K. were  supported by \emph{Funda\c{c}\~{a}o de Amparo
\`{a} Pesquisa do Estado de S\~{a}o Paulo (FAPESP)}, Brazil.
\end{acknowledgments}

\end{document}